\documentclass[aps,prl,twocolumn,showpacs,preprintnumbers]{revtex4}
\usepackage{amsmath}
\usepackage{graphicx}
\usepackage{dcolumn}
\usepackage{bm}
\usepackage{subfigure}
\usepackage{amsfonts}
\usepackage{amssymb}

\begin{document}

\title{\textbf{On the Boltzmann-Grad Limit for the classical hard-spheres
system}}
\author{Massimo Tessarotto\thanks{%
Electronic-mail: M.Tessarotto@cmfd.univ.trieste.it}}
\affiliation{$^{1}$Dipartimento di Matematica e Informatica, Universit\`{a} di Trieste,
Italy\\
$^{2}$Consorzio di Magnetofluidodinamica, Trieste, Italy\\
}
\date{\today }

\begin{abstract}
Despite the progress achieved by kinetic theory, the search of possible
exact kinetic equations remains elusive to date. This concerns,
specifically, the issue of the validity of the conjecture proposed by Grad
(Grad, 1972) and developed in a seminal work by Lanford (Lanford, 1974) that
kinetic equations - such as the Boltzmann equation for a gas of classical
hard spheres - might result exact in an appropriate asymptotic limit,
usually denoted as Boltzmann-Grad limit. The Lanford conjecture has actually
had a profound influence on the scientific community, giving rise to a whole
line of original research in kinetic theory and mathematical physics.
Nevertheless, certain aspects of the theory remain to be addressed and
clarified. The purpose of this paper is to investigate the possible
existence of the strong Boltzmann-Grad limit for the BBGKY hierarchy. \
Contrary to previous approaches in which the w*-convergence was considered
for the definition of the Boltzmann-Grad limit functions, based on their
construction in terms of time-series expansions obtained from the BBGKY
hierarchy, here we look for the possible existence of strong limit functions
in the sense of local convergence in phase space. \ The result is based on
the adoption of the Klimontovich approach to statistical mechanics,
permitting the explicit representation of the $s$-body reduced distribution
functions in terms of the Klimontovich probability density.
\end{abstract}

\maketitle


\section{1 - Introduction - \ Foundations of CKT}

Basic issues concerning the foundations of classical kinetic theory (CKT)
still remain unanswered. Since the criticism raised by Zermelo \cite%
{Zermelo1896} on Boltzmann equation \cite{Boltzmann1872} the possibility of
a rigorous construction of kinetic equations for classical gases has been
the subject of investigations by many. In fact it is well known that
Boltzmann himself obtained his famous equation using only physically
plausible arguments, not first principles, i.e., the microscopic dynamics of
the hard-sphere system. Despite the progress achieved by CKT in the last
decades, its difficulty is notorious and is associated with the asymptotic
character of kinetic equations, which potentially makes it hard - or even
impossible - the construction of exact kinetic equations for many systems,
in particular for classical systems of interacting hard spheres.\ As such,
the investigation of the rigorous of CKT represents a challenge both for
mathematical analysis and mathematical physics alike. Its importance for gas
dynamics and classical hydrodynamics goes beyond the academic interest. \
One such problem is posed by the possible existence of exact kinetic
equations obtained in the so-called Boltzmann-Grad (B-G) limit, which should
apply to infinite classical systems of interacting particles, provided the
microscopic phase-space distribution function (PSDF) satisfies physical
constraints and functional settings to be properly defined. The goal of this
investigation is to propose a novel approach to CKT, pertaining to
hard-sphere systems, based on the investigation of the properties of the
so-called limit functions which are obtained for such systems in the B-G
limit. In particular, we wish to investigate the limit functions which enter
the BBGKY and prove that - contrary to common belief - they do not generally
belong to the functional class of the solutions of the asymptotic Boltzmann
hierarchy. This means, in particular, that in the case of smooth and hard
sphere systems, the limit functions do not generally satisfy the exact
Boltzmann kinetic equation, although the explicit construction of asymptotic
solutions for the BBGKY hierarchy can be achieved based on the determination
of suitable weakly-convergent sequences.

\subsection{1a - Basic motivations: 'ab initio' approaches}

Classical statistical mechanics, and in particular kinetic theory
represents, is a sense, one of the unsolved problems of classical mechanics.
In fact, although the microscopic statistical description (MSD) of classical
dynamical systems formed by $N$-body systems is well known, a complete
knowledge of their solutions is generally not achievable. From the
mathematical viewpoint it provides an example of axiomatic approach
following from first principles and as such it must be considered as an 'ab
initio' formulation. Two equivalent treatments of MSD are known, which are
based respectively on the introduction of a phase-space distribution
function (PSDF) either on the $N$-body phase-space $\Gamma _{N}$ or,
respectively, on the 1-particle phase-space $\Gamma _{1}.$ In the $\Gamma
_{N}$ -approach the PSDF is the so-called microscopic PSDF $f_{N}$. It
follows that $f_{N}$ obeys the Liouville equation, whose characteristics are
simply the phase-space trajectories of the same dynamical system, to be
identified with a classical $N$-body system \cite{Grad1958,Cercignani1969}.
This equation is equivalent to a hierarchy of equations (the so-called BBGKY
hierarchy) for a suitable set of $s$-particles distribution functions ($%
f_{s}^{(N)}$), obtained letting $s=1,..,N-1$, which are uniquely related to
the corresponding PSDF. On the other hand, in the $\Gamma _{1}-$approach the
PSDF (the Klimontovich probability density $k^{(N)}$, defined in the $\Gamma
_{1}-$space) evolves in time by means of the Klimontovich equation \cite%
{Klimontovich}. Also for this equation the characteristics are just the
phase-space trajectories of the $N-$body system, this time - however -
projected on the $\Gamma _{1}-$space. Therefore, in both cases it is
actually necessary to determine the phase-space trajectories of all the
particle. Hence, for classical systems characterized by a large number of
particles ($N\gg 1$), the computational complexity (of this problem) is
expected to prevent, in general, any direct calculation of the
time-evolution either of the $N$-body or any of the the s-body
distributions. This has justified the constant efforts placed so-far for the
search of 'reduced' statistical descriptions, of which kinetic theory (KT)
is just an example. This is intended in order to achieve efficient
statistical descriptions especially suitable for complex dynamical systems,
including both gases and plasmas. \ Precisely, the primary goal of KT is the
search of statistical descriptions, either exact or in some sense
approximate, whereby the whole dynamical system is associated only to the
one-particle kinetic distribution function ($f_{1}$) defined on the
one-particle phase-space $\Gamma _{1}$, without requiring the knowledge of
the dynamics of the whole dynamical system. As a consequence in
KT-descriptions the evolution equation of the kinetic distribution function,
to be denoted as kinetic equation, is necessarily assumed to depend
functionally, in some suitable sense, only on the same distribution function
and the one-particle dynamics. In particular, one of the most successful
developments of KT is doubtless related to the so-called 'ab initio'
approaches. These are to be intended (in contrast to heuristic or model
equations) as the KT's which are obtained deductively - by suitable
approximation schemes and assumptions - from the corresponding exact MSD. In
traditional approaches usually KT is obtained adopting the $\Gamma _{N}$
-approach to MSD \cite{Grad1958,Cercignani1969,Cercignani1975}. However,
also the Klimontovich method (based on the $\Gamma _{1}-$approach) can be
used \cite{Klimontovich}, since it is completely equivalent to that based on
the $\Gamma _{N}$ -approach \cite{Pin2001}. In all cases KT's have the goal
of determining the evolution of suitable fluid fields,\emph{\ }associated to
prescribed fluids,\emph{\ }which are expressed as \emph{velocity moments of
the kinetic distribution function } $f_{1}$ \emph{\ }and satisfy an
appropriate set of fluid equations, generally not closed, which follow from
the relevant kinetic equation. 'Ab initio' kinetic theories are - however -
usually asymptotic in character. Namely, kinetic equations are typically
satisfied only in an approximate (and asymptotic) sense and in a finite time
interval, under suitable assumptions. These require in particular that $%
f_{N} $ (and all $f_{s},$ for $s=1,N-1$) must belong to a suitable
functional class (here denoted as $\left\{ f_{N}\right\} _{I}$) so that $%
f_{N}$ as well as the related $s-$particle distribution functions $%
f_{s}^{(N)}$ satisfy appropriate initial and boundary conditions.

\subsection{1b - Asymptotic kinetic theories}

A well-known asymptotic kinetic equation of this type is provided by the
Boltzmann kinetic equation for a classical gas formed by $N$ smooth rigid
spheres of diameter $d$ (Grad,1958 \cite{Grad1958}), which is obtained from
the exact equation of the BBGKY hierarchy for the one-particle kinetic
distribution, i.e.,
\begin{equation}
{\ {F}_{1}(\mathbf{x}_{1},t)f}_{1}^{(N)}{(\mathbf{x}_{1},t)=d}^{2}{(N-1)}%
\left\{ {C_{1}f}_{2}^{(N)}\right\} _{{(\mathbf{x}_{1},t)}},  \label{Eq.1}
\end{equation}%
where $F_{1}$ and ${C_{1}f}_{2}^{(N)}$ are respectively the free-streaming
operator${\ F}_{1}(\mathbf{x}_{1},t){=}\frac{\partial }{\partial t}{+}%
\mathbf{v}_{1}{\cdot }\frac{\partial }{\partial \mathbf{r}_{1}}$ and the
BBGKY collision operator ${C}_{1}{\rho }_{2}^{(N)}{=}\int {\ d}\mathbf{v}_{2}%
{d\Sigma }_{12}{\rho }_{2}^{(N)}\mathbf{v}_{12}\mathbf{\ \cdot n}_{12.},$
containing the integration on velocity and the solid angle ${d\Sigma }_{12}$
of particle $2,$ and the notation is standard \cite%
{Grad1958,Cercignani1969,Cercignani1975}. \ Thus ${\mathbf{x}_{1}=(\mathbf{r}%
_{1},\mathbf{v}_{1})}$ is the Newtonian state of particle $1,\mathbf{n}%
_{12.} $ is unit vector $\mathbf{n}_{12.}={\mathbf{r}_{12}}/\left\vert {%
\mathbf{r}_{12}}\right\vert $, \ while $\mathbf{v}_{12}={\mathbf{v}_{1}-%
\mathbf{v}_{2}}\mathbf{\ }$and $\mathbf{r}_{12.}={\mathbf{r}_{1}-\mathbf{r}%
_{2}}$ are respectively the relative velocities and position vectors of
particles $1$ and $2$. \ For definiteness, in the remainder we adopt a
dimensionless notation whereby all relevant functions (in particular, the
Newtonian particle state ${\mathbf{x}_{1}=(\mathbf{r}_{1},\mathbf{v}_{1}),}$
the time ${t}$ ${,}$ the particle diameter $d$ and the volume of the
configuration space $V$) are considered non-dimensional. \ Eq. (\ref{Eq.1})
can also be written in the integral form%
\begin{eqnarray}
&&\left. {f}_{1}^{(N)}({\mathbf{x}_{1}(t),t)=f}_{1}^{(N)}({\mathbf{x}_{1o},t}%
_{o}{)+}\right.  \label{Eq.1b} \\
&&\left. {+d}^{2}{(N-1)}\int\limits_{t_{o}}^{t}dt^{\prime }{\left\{ {C_{1}f}%
_{2}^{(N)}\right\} _{{(\mathbf{x}_{1}(t}^{\prime }){,t}^{\prime }{)}},}%
\right.  \notag
\end{eqnarray}%
by integrating the previous equation along the Lagrangian characteristics ${%
\mathbf{x}_{1}(t)}$. The equation can be iterated by representing in a
similar way the $s-$particle joint probability densities ${f}_{s}^{(N)}$ for
$s=2,3,{..}$ etc., obtained integrating the corresponding equations of the
BBGKY hierarchy. The transition from the $1$-particle equation (\ref{Eq.1})
can be obtained by adopting a suitable asymptotic approximation and
appropriate assumptions on the joint probability densities\cite%
{Grad1958,Grad1972}. \ These require, in particular, the introduction of the
so-called\textit{\ rarefied gas ordering} (\textit{RG ordering}) for the
relevant physical parameters, to be intended both in a global and local
sense. \ More precisely, by imposing that $\varepsilon =1/N$ is an
infinitesimal, the particle diameter $d$, the volume $V$ of the
configuration space ($\Omega $) and the particle mass $m$ must be suitably
ordered in terms of $\varepsilon .$ Thus, the \textit{global ordering} is
obtained requiring that $d$ and $m$ are respectively infinitesimals of order
$\varepsilon ^{1/2}$ and $\varepsilon ,$ whereas the volume of the
configuration space is taken of order $O(1)$ (Grad,1958 \cite{Grad1958}).
This implies that average volume fraction $\overline{\overline{\eta }}\equiv
4\pi Nd^{3}/3V$ results necessarily an infinitesimal of order $\varepsilon
^{1/2}$. In addition, to assure that the gas is rarefied everywhere in $%
\Omega ,$ the local volume fraction \textit{\ }$\eta (\mathbf{r,}t)\equiv
4\pi n(\mathbf{r,}t)d^{3}/3V$\textit{\ }$\ $must be assumed of order $%
\varepsilon ^{1/2}$ \emph{everywhere} in $\Omega \times I_{o1}$ (\textit{%
local ordering})$.$ Here\ $I_{o1}$ is the time interval $I_{o1}=\left[
t_{o},t_{1}\right] ,$ with $\Delta t=t_{1}-t_{o}$ defined so that $\Delta
t\sim O(1)$ and $n(\mathbf{r,}t)$ is the local number density. Thus, the
local ordering prevents the number density from becoming so large for $\eta (%
\mathbf{r,}t)$ to be locally finite, i.e., of order $O(1).$ It is
well-known, in fact, that if $\eta (\mathbf{r,}t)$ becomes locally of order $%
O(1),$ particle correlations (in particular two-particle correlations) may
become non-negligible also on a larger scale \cite%
{Grad1958,Grad1967,Tsuge1970}. These correlations, which are not generally
expected to decay rapidly in time \cite{Grad1967}, can be long-range in
character \cite{Piasecki2007}. \ Instead, in validity of the RG ordering \ -
and in particular imposing of the local ordering indicated above - the
following conditions are assumed to be satisfied uniformly in phase-space
and at least in a finite time interval $I=\left[ t_{o},t_{1}\right] ,$ with $%
\Delta t=t_{1}-t_{o}$ such that $\Delta t\sim O(1)$:

\begin{itemize}
\item Assumption \#1 - in $\Gamma _{s}\times I_{o1}$, the approximate (i.e.,
asymptotic) joint probability densities $f_{s}(\varepsilon )$ (for any $s\in
\mathbb{N}$ \ with $s\ll N$)\ are smooth and bounded ordinary functions
defined in $\Gamma _{s}\times I_{o1},$ where $\Gamma _{s}$ is the $s$%
-particle phase-space.

\item Assumption \#2 - the \textit{asymptotic factorization condition }(AFC)
\begin{eqnarray}
&&\left. f_{s}(\varepsilon ,\mathbf{x}_{1}\mathbf{,..x}_{s},t)=\prod%
\limits_{i=1,s}f_{1}(\varepsilon ,\mathbf{x}_{i},t)\left[ 1+\right. \right.
\label{AFC} \\
&&\left. +\Theta (t-t_{o})o(\varepsilon ^{\alpha })\right]  \notag
\end{eqnarray}%
is satisfied identically for any $s\in
\mathbb{N}
$ such that $s/N$ is an infinitesimal of order $\varepsilon $. Here $f_{1}$ $%
(\varepsilon ,\mathbf{x}_{i},t)$ (for $i=1,s$) \ is the one-particle
probability density which satisfies the asymptotic Boltzmann equation
\begin{equation}
{\ {F}_{1}(\mathbf{r}_{1},\mathbf{v}_{1},t)\ f}_{1}(\varepsilon ,){=d}^{2}{\
NC_{1}f}_{2}(\varepsilon ,),  \label{asymptotic Boltzmann eq}
\end{equation}
\end{itemize}

and $\Theta (t-t_{o})$ is the Heaviside theta function which vanishes for $%
t=t_{o}.$

If the RG ordering and the previous assumptions hold locally (i.e., in the
whole phase-space $\ \Gamma _{1}$ and at least in a finite time interval $%
I_{o1}\equiv \left[ t_{o},t_{1}\right] $), \ the Boltzmann equation (\ref%
{asymptotic Boltzmann eq}) is expected to be locally valid in the same
domain \cite{Shinbrot1984,Illner1986,Pulvirenti1987} at least in an
asymptotic sense.

\ Even if the rigorous proof of the global validity of the Boltzmann
equation for arbitrary initial and boundary conditions has yet to be
reached, its success in providing extremely accurate predictions for the
dynamics of rarefied gases and plasmas is well known (see for example,
Cercignani, 1969 \cite{Cercignani1969}; Frieman, 1974 \cite{Frieman1975}).

\subsection{1c - The Boltzmann-Grad limit and the Lanford conjecture}

Nevertheless, basic issues remain to be clarified regarding the rigorous
theoretical foundations of kinetic theory. Following the conjecture
suggested originally by Grad (Grad, 1972 \cite{Grad1972}), it is generally
believed that in certain "singular limits" the kinetic equations - such as
the Boltzmann equation for a gas of classical hard spheres - may result
exact. A basic difficulty is to properly formulate the related mathematical
problem and to ascertain in a rigorous way the possible validity of such a
type of statements. \ One such problem refers in particular to the search of
possible exact kinetic equations and, specifically, the conjecture (here
denoted as \emph{Lanford conjecture}) proposed by Lanford in a seminal paper
(Lanford, 1974 \cite{Lanford1975a}; see also Grad, 1972 \cite{Grad1972} and
Frieman, 1974 \cite{Frieman1975}),\ that the Boltzmann kinetic equation for
a gas of classical hard spheres might result exact in an appropriate
asymptotic limit, denoted as \textit{Boltzmann-Grad} (B-G) \textit{limit}.

The B-G limit is customarily intended as the limiting "regime" where the
total number of particles $N$ goes to infinity, while the
configuration-space volume $V$ remains constant, the particle diameter $d$
goes to zero in such a way that $Nd^{2}$ approaches a finite non-zero
constant and the average mass density $Nm/V=M/V$ remains finite (Grad, 1972
\cite{Grad1972}; Lanford, 1974 \cite{Lanford1975a}), i.e., there results:
\begin{eqnarray}
&&\left. \frac{1}{N},d,m\rightarrow 0,\right.  \notag \\
&&\left. \frac{Nd^{2}}{V}\rightarrow k_{1},\right.  \label{B-G-1} \\
&&\left. M=\frac{mN}{V}\rightarrow k_{2},\right.  \notag
\end{eqnarray}%
where $k_{i}$ (for $i=1,2$) are prescribed non-vanishing positive and finite
constants. In the case of plasmas further analogous requirements are placed
on the total electric charge and current carried by each particle species
\cite{Frieman1975,Tessarotto1999a}. In the original Lanford formulation, it
was conjectured that, subject to suitable initial and regularity conditions,
the one-particle probability density determined by the integral equation (%
\ref{Eq.1b}) converges weakly, in the sense of weak * convergence, to a
limit function $f_{w1}(\mathbf{x}_{1},t)\equiv L_{w}^{\ast }f_{1}^{(N)}(%
\mathbf{x}_{1},t),$ $L_{w}^{\ast }$ denoting an appropriate operator, to be
denoted as \emph{weak B-G limit operator} and\emph{\ }$f_{w1}(\mathbf{x}%
_{1},t)$ the solution of the equation [stemming from Eq.(\ref{Eq.1b})]\
\begin{eqnarray}
&&\left. {f}_{w1}({\mathbf{x}_{1}(t),t)=f}_{w1}({\mathbf{x}_{1o},t}_{o}{)+}%
\right.  \label{time-series solution} \\
&&+{Vk}_{1}\int\limits_{t_{o}}^{t}dt^{\prime }L_{w}^{\ast }{\left\{ {C_{1}f}%
_{2}^{(N)}\right\} _{{(\mathbf{x}_{1}(t}^{\prime }){,t}^{\prime }{)}}.}
\notag
\end{eqnarray}%
In this meaning the conjecture was actually proven true by Lanford, at least
in a partial sense, namely for a time interval which has an amplitude not
exceeding one fifth of the mean free path measured from an initial time%
\textsl{\ }$t_{o}$ (\emph{Lanford Theorem}). \ The proof, \ first presented
in his work on the B-G limit (Lanford, 1974 \cite{Lanford1975a}) under the
assumption of factorization at the initial time (i.e., Eq.(\ref{AFC}) taken
at $t=t_{o},$ while letting $\varepsilon \rightarrow 0$), was actually
reached by proving the convergence, in the sense of weak *-convergence, of
the limit time-series solution (\ref{time-series solution}). Obviously, this
result does not suffice to justify possible meaningful physical
applications. Nevertheless, the conjecture has actually had a profound
influence on the scientific community, giving rise to a whole line of
original research in kinetic theory and mathematical physics. In particular,
the proof has been extended to more general situations \cite%
{Shinbrot1984,Illner1986,Pulvirenti1987,Gerasimenko1990}. Nevertheless,
despite the progress achieved by kinetic theory, the issue of existence of
the B-G limit remains, however, open to date. Despite the significant number
of theoretical papers appeared in the literature in the last three decades,
the issue of the validity of the Boltzmann equation in the B-G limit \cite%
{Lanford1975a,Alexander1975,Lanford1976,Kaniel1978,Van
Beijeren1980,Spohn1981,Lanford1981,Spohn1983,Shinbrot1984,Spohn1984,Illner1986,Pulvirenti1987,Illner1987b,Gerasimenko1990,Esposito1989,Caprino1996}
is probably the one for which a complete understanding is not yet available.
Several aspects of the theory remain to be addressed and clarified. In fact,
it is strongly doubtful whether the Boltzmann equation can apply for
arbitrary times and for general situations. Regarding this issue a general
remark must be made. Just as what happens for the RG ordering (see previous
discussion on the RG ordering), the global conditions defined by Eqs. \ref%
{B-G-1} are generally insufficient to specify uniquely the B-G limit. It is
obvious, in fact, that in principle the B-G limit may be taken locally in
arbitrary ways, so that it is generally insufficient to permit the validity
of a kinetic description which requires the complete absence (or the
neglect) of binary or higher-order particle correlations \cite{Grad1972}.
This means that to assure the local validity of the Boltzmann equation, the
B-G limit should be intended also in a local sense, by adding some
additional appropriate prescription. For example, the definition (of the B-G
limit) might be intended, for example, in the sense of van Hove. For this
purpose let us consider a sequence of bounded open regions $\Omega (r)$ all
included in the configuration domain $\Omega \equiv \Omega (D)\subset
\mathbb{R}^{3},$ with $0<r\leq D$ a real parameter (in particular, if $%
\Omega (r)$ and $\Omega (D)$ are identified with spherical domains, with $%
\Omega (r)$ $\subseteq $ $\Omega (D),$ $r,D$ are the corresponding radii).
In addition let us denote by $V(r)$ the volume of the set of points included
in $\Omega (r)$ and with $V(D)$ the volume of the entire region $\Omega (D).$
Then the B-G limit can be intended, for example, in the sense that locally,
besides the global limit conditions defined above (\ref{B-G-1}), there
results for any $0<r\leq D:$

\begin{eqnarray}
&&\left. \frac{N(r)}{N(D)}\sim \frac{M(r)}{M(D)}\right.  \label{B-G-2} \\
&&\left. \frac{N(r)d(N(D))^{2}}{V(r)}\sim \frac{N(D)d(N(D))^{2}}{V(D)}%
\rightarrow const\right.  \label{B-G-3}
\end{eqnarray}%
where $N(r)$ and $N(\Omega )$ are respectively the number of particles in
the regions of volumes $V(r)$ and $V(\Omega )$ and again the constant is
assumed finite and non-vanishing.

\subsection{1d - Goals of the investigation}

However, even imposing the additional local conditions \ref{B-G-2} and \ref%
{B-G-3}, the issue of the validity of the Lanford conjecture remains
unsettled when the B-G limit is meant in the sense of strong (i.e., uniform)
convergence for the sequences\ for the joint-probability densities $\left\{
f_{s}^{(N)}\right\} .$ The reason is that the limit functions of the
sequences $\left\{ f_{s}^{(N)}\right\} ,$ to be defined in the sense strong
convergence for the B-G limit, do not belong necessarily to the same
functional class of the same sequence. As a result, while weak convergence
in the sense indicated above is in principle still warranted, it might still
occur, in particular, that the (strong) B-G $1$-particle limit function is
not a solution of the Boltzmann equation.

Here we want to investigate a basic issue - preliminary w.r. to the
treatment of \ the Boltzmann equation - namely the validity of the Lanford
conjecture for the BBGKY hierarchy itself, to be intended in the sense of
the strong B-G limit, here denoted as \emph{strong Lanford conjecture}. \
The conjecture requires, that there exists a \emph{strong B-G limit operator}
$L^{\ast }$ which, applied to the equations of the BBGKY hierarchy for the $%
s-$particle joint probability densities ${f}_{s}^{(N)}$, delivers the
corresponding equations of the Boltzmann hierarchy for the corresponding
limit functions $\ f_{s}=L^{\ast }f_{s}^{(N)}$ that a suitable
limit-hierarchy must exist (to be denoted as the Boltzmann hierarchy) for
suitable $s$-particle limit functions $f_{s}\equiv L^{\ast }f_{s}^{(N)}$
(for $s\in \mathbb{N}$). In particular in the case of the BBGKY equation for
one-particle probability density
\begin{equation}
{\ F}_{{\ 1}}{\ f}_{1}^{(N)}{\ =d}^{2}({N-1)C_{1}f}_{2}^{(N)},
\label{eq. for f1^N}
\end{equation}%
applying the operator $L^{\ast }$ to both sides it should result identically
the (exact) equation of the Boltzmann hierarchy
\begin{equation}
{\ F}_{{\ 1}}{\ f}_{1}{\ =d}^{2}{NC_{1}f}_{2}.  \label{limit equation}
\end{equation}%
This means that the strong limit functions $f_{s}=L^{\ast }f_{s}^{(N)}$ (for
$s=1,2$) should have the property that:

\begin{itemize}
\item a) the limit function $\ f_{1}=L^{\ast }f_{1}^{(N)}$ should belong to
the functional class of the solutions of the Boltzmann hierarchy;

\item b) it should result identically
\begin{equation}
{\ }\left[ {{L^{\ast },F}_{{\ 1}}}\right] {{\ }f}_{1}^{(N)}(\mathbf{x}%
_{1},t)\equiv 0,  \label{commutation rule}
\end{equation}%
where $\left[ {{L^{\ast },F}_{{\ 1}}}\right] $ denotes the commutator $\left[
{{L^{\ast },F}_{{\ 1}}}\right] ={{L^{\ast }F}_{{\ 1}}-{F}_{{\ 1}}{L^{\ast }.}%
}$ This means that the operators $L^{\ast }$ and ${{F}_{{\ 1}}}$ should
commute when acting on the one-particle probability density;

\item c) and finally the following limit: $\ $%
\begin{equation}
{{L^{\ast }}d}^{2}{\ (N-1)C_{1}f}_{2}^{(N)}={d}^{2}{\ NC_{1}{L^{\ast }}f}%
_{2}^{(N)}  \label{limit collision operator}
\end{equation}%
should hold.
\end{itemize}

Main goal of the investigation is to analyze the possible validity of the
strong Lanford conjecture here proposed and in particular whether properties
a)-c) are generally fulfilled or not, in other words, whether \emph{the
strong B-G limit function }$\ f_{1}=L^{\ast }f_{1}^{(N)}$\emph{\ may belong
or not to the functional class of the solutions of the corresponding
equation of the Boltzmann hierarchy}.

The possible solution of this problem goes beyond the academic interest. In
fact, not only it represents a difficult theoretical problem, but it is
related to the very foundations of statistical mechanics. As such, its
investigation represents a challenge both for mathematical analysis and for
theoretical physics. The possible solution of the riddle posed by the strong
Lanford conjecture provides, in fact, a new interesting starting point for
theoretical research in kinetic theory. \ This paper will analyze for this
purpose the classical model based on a gas of hard-smooth spheres. The
approach is based on the adoption of the Klimontovich approach to
statistical mechanics, permitting the explicit representation of the $s$%
-body reduced distribution functions in term of the Klimontovich probability
density.

\section{2 - MSD approach for the hard-sphere system in the $\Gamma _{1}-$%
phase-space}

Let us consider the time evolution of a system ($S_{N}$) of $N$ identical
smooth spheres. The particles are assumed of diameter $d,$ mass $m$ and
immersed in a compact connected configuration domain $\Omega \subset
\mathbb{R}
^{3},$ with prescribed fixed boundary $\partial \Omega $ represented by a
smooth regular surface \cite{Grad1958}. In the sequel particles are assumed
to be subject only to binary and unary elastic collisions. Both occur when
the boundaries of the particles and/or $\partial \Omega $ come into mutual
contact in such a way that the colliding boundaries, before collision, have
a non-vanishing relative velocity. Multiple collisions - i.e., simultaneous
collisions between particles and/or $\partial \Omega $, by assumption, are
considered as\ sequences of binary and/or unary collisions. \ For
definiteness, we shall assume all particles to be 'hard', i.e., such that
their boundaries are rigid and furthermore that each particle can come into
contact with $\partial \Omega $ only in a single point. This condition is
satisfied, for example, if $\partial \Omega $ is identified with a spherical
surface (of radius $R_{o}$)$.$ In such a case only a subset of admissible
configurations of $\Omega $ is actually permitted$.$ This is defined as the
set $\overline{\Omega }=\left\{ \mathbf{r:r\in \Omega ,}\overline{\Theta }%
_{i}(\mathbf{r,\mathbf{\xi }}(t),t)=1,\forall i=1,N\right\} ,$ where $%
\overline{\Theta }_{i}(\mathbf{r,\mathbf{\xi }}(t)\mathbf{,}t)$ is the
occupation function for the $i$-th particle, $\overline{\Theta }_{i}(\mathbf{%
r,\mathbf{\xi }}(t),t)\equiv 1-\sum_{\substack{ j=1,N \\ i\neq j}}\overline{%
\Theta }(d-\left\vert \mathbf{r-r}_{j}(t)\right\vert )$ $-\overline{\Theta }%
(d/2-\left\vert \mathbf{r-r}_{W}\right\vert ).$ Here $\mathbf{r}_{W}$ is a
position vector defining an arbitrary point of the boundary $\partial \Omega
,$ $\mathbf{\mathbf{\xi }}(t)$ denotes the $N-$particle configuration vector
$\mathbf{\mathbf{\xi }}(t)\equiv \left\{ \mathbf{\mathbf{r}}_{1}(t),..,%
\mathbf{\mathbf{r}}_{N}(t)\right\} ,$ while $\overline{\Theta }(x)$ is the
so-called strong Heaviside step function i.e., $\overline{\Theta }(x)=1,0$
if $x>0,$ $x\leq 0$. \ We stress that in the definition of all the
occupation functions (both $\overline{\Theta }_{i}$ and $\overline{\Theta }%
_{i}$ given below) the configuration vector $\mathbf{\mathbf{\xi }}(t)$ is
defined in such a way that the position vectors $\mathbf{\mathbf{r}}%
_{1}(t),..,\mathbf{\mathbf{r}}_{N}(t)$ are always considered mutually
admissible. This means in particular that for all $i,j=1,N$ (with $i\neq j$)
it must result in $\left\vert \mathbf{\mathbf{r}}_{i}(t)-\mathbf{\mathbf{r}}%
_{j}(t)\right\vert \geq d.$ \ One can define in a similar way also the
subset of $\Omega $ (to be denoted as $\widehat{\Omega }$) in which no
interactions occur (for all particle of $S_{N}$) as well as the
corresponding occupation function,\ to be denoted as strong occupation
function$.$ The latter reads for the $i$-th particle:
\begin{eqnarray}
\Theta _{i}(\mathbf{r,\xi }(t)\mathbf{,}t) &\equiv &1-\sum_{\substack{ j=1,N
\\ i\neq j}}\Theta (d-\left\vert \mathbf{r-r}_{j}(t)\right\vert )-
\label{occupation number} \\
&&-\Theta (d/2-\left\vert \mathbf{r-r}_{W}\right\vert ),  \notag
\end{eqnarray}%
where $\Theta (x)$ is the strong Heaviside step function i.e., $\Theta
(x)=1,0$ if $x\geq 0,$ $x<0.$ It follows that the set $\widehat{\Omega }$ is
simply the subset of $\Omega $ in which the equations $\Theta _{i}(\mathbf{r,%
\mathbf{\xi }}(t),t)=1$ are satisfied identically for all particles, i.e.,
for all $i=1,N$. \ Moreover, by assumption, particles are 'smooth'. This
means that they undergo only interactions (collisions) which conserve the
angular momenta of all particles. Hence, the state of $S_{N}$ is uniquely
defined by ensemble of states $\mathbf{x}(t)=\left\{ \mathbf{x}_{1}(t),...%
\mathbf{x}_{N}(t)\right\} \equiv (\mathbf{\mathbf{\xi }}(t),\mathbf{\mathbf{%
\eta }}(t)),$ where $\mathbf{x}_{i}(t)$ (for $i=1,N$) \ represents the state
of each particle defined by the vector $\mathbf{x}_{i}(t)=\left\{ \mathbf{r}%
_{i}(t),\mathbf{v}_{i}(t)\right\} ,$ $\mathbf{r}_{i}(t)$ and $\mathbf{v}%
_{i}(t)$ denoting the positions and velocities of the centers of each
sphere. Thus,\ $\mathbf{\mathbf{\eta }}(t)\equiv \left\{ \mathbf{\mathbf{v}}%
_{1}(t),..,\mathbf{\mathbf{v}}_{N}(t)\right\} $ while each vector $\mathbf{x}%
_{i}(t)$ (for $i=1,N$) spans the\ one-particle admissible phase-space $%
\Gamma _{1(i)}=\overline{\Omega }\times
\mathbb{R}
^{3}.$ \ We notice that in a similar way it is possible to define admissible
and forbidden sub-domains in the $N$-particle configuration-space $\Omega
^{N}$ and in the corresponding phase-space $\Gamma _{N}$ $=\Omega ^{N}\times
\mathbb{R}
^{3N}.$\ In particular, we denote by $\overline{\Gamma }_{N}$ (respectively $%
\overline{\Gamma }_{s}$) the admissible subsets of $\Gamma _{N}$ ($\Gamma
_{s}$) in which the configurations of all $N$ particles (respectively of the
first $s$ particles) are all admissible and $\Gamma _{N}^{\ast }=\Gamma _{N}-%
\overline{\Gamma }_{N}$ ($\Gamma _{s}^{\ast }=\Gamma _{s}-\overline{\Gamma }%
_{s}$) its complementary set, denoting the forbidden sub-domain of $\Gamma
_{N}$ ($\Gamma _{s}$).

Regarding particle dynamics, the motion of each ($i$-th) particle of $S_{N}$
is assumed inertial in any open subset \ $\left] t_{k,}^{(i)}t_{k+1}^{(i)}%
\right[ $ of $I$ not containing collision events for the same particle (the
time interval between two successive collision events occurring). Finally,
at an arbitrary collision time for the same particle $(t_{c}^{(i)}),$ the
phase-flow is defined respectively, for binary and unary interactions, by
the elastic two- and one-particle collision laws\cite{Grad1958}, which
uniquely relate its states before [$\mathbf{x}_{i}^{-}(t_{c})$] and after
collision [$\mathbf{x}_{i}^{+}(t_{c})$]. \ As a consequence, the mapping
provided by the phase-flow between an arbitrary admissible initial state $%
\mathbf{x}(t_{o})=\mathbf{x}_{o},$ with $\mathbf{x}_{o}=\left\{ \mathbf{x}%
_{1o},..,\mathbf{x}_{No}\right\} ,$ and its image at an arbitrary time $t\in
I\equiv
\mathbb{R}
,$ $\mathbf{x}(t)=\chi (\mathbf{x}_{o},t-t_{o})\in \overline{\Gamma }_{N}$
is manifestly defined globally in $\Gamma _{N}$ $\times I$. \

The microscopic statistical description of $S_{N}$ adopting the $\Gamma _{N}$%
-phase-space description - \ and based on the introduction of the PSPD $%
f_{N}(\mathbf{x,}t)$ in $\Gamma _{N}$ - is well-known \cite%
{Grad1958,Cercignani1969,Cercignani1975}. The relevant mathematical
framework is recalled in the Appendix (see in particular Theorem 1).

The MSD for $S_{N}$ on the phase-space $\Gamma _{1}$ can, instead, be
achieved by introducing the Klimontovich probability density for $S_{N}$ on
the same phase-space. Following the Klimontovich approach\cite{Klimontovich}%
,\ \ this is defined as a probability distribution on $\Gamma _{1}$ which is
assumed as non-vanishing only along the subsets of the trajectories of the
particles of $S_{N}$ system (all mapped on the phase-space $\Gamma
_{1}\equiv \Omega \times
\mathbb{R}
^{3}$) where all particles of $S_{N}$ are not subject to interactions, i.e.,
in the subset $\widehat{\Gamma }_{1}=$ $\widehat{\Omega }\times
\mathbb{R}
^{3}$. Hence, the Klimontovich probability density necessarily takes the
form:

\begin{equation}
k^{(N)}(\mathbf{y,}t)=\frac{1}{N}\sum_{i=1,N}\delta (\mathbf{y-x}%
_{i}(t))\Theta _{i}(\mathbf{r,\mathbf{\xi }}(t),t)
\label{Klimontovich density}
\end{equation}%
where $\Theta _{i}(\mathbf{r,}t)$ is the occupation function \ defined by
Eq.(\ref{occupation number}) and $\mathbf{y=}\left( \mathbf{r,v}\right) \in
\Gamma _{1}$. Hence, it follows that $k^{(N)}(\mathbf{y,}t)$ in $\widehat{%
\Gamma }_{1}$ satisfies identically the $\Gamma _{1}$-space Liouville
equation:%
\begin{equation}
\left( \frac{\partial }{\partial t}+\mathbf{v}\cdot \nabla \right) k^{(N)}(%
\mathbf{y,}t)=0.  \label{Gamma-1-Liouville equation}
\end{equation}%
Then the following theorem applies:

\bigskip

\textbf{Theorem 2 - }$\Gamma _{1}$-\textbf{MSD for} $S_{N}$

\emph{Let us assume that for }$S_{N}$\emph{\ the microscopic probability
density } $f_{N}(\mathbf{x,}t)$ \emph{satisfies assumptions of THM.1 (see
Appendix). Then it follows that:}

\emph{A) in }$\Gamma _{1}\times I$ \emph{(for any }$\mathbf{y\equiv (r,v)\in
\Gamma }_{1}$ \emph{and} $t\in I\subseteq
\mathbb{R}
$\emph{)} \emph{the 1-particle probability density }$f_{1}^{(N)}(\mathbf{y}%
,t)$ \emph{admits} \emph{the integral representation in terms of the initial
microscopic probability density }$f_{N}(\mathbf{x}_{o}\mathbf{,}t_{o})$\emph{%
\ :}

\begin{eqnarray}
&&\left. f_{1}^{(N)}(\mathbf{y},t)=\int\limits_{\Gamma _{N}}d\mathbf{x}%
_{o}f_{N}(\mathbf{x}_{o}\mathbf{,}t_{o})\frac{1}{N}\right.
\label{one-particle PFD-dependence-initial-condition} \\
&&\sum_{i=1,N}\delta (\mathbf{y-\chi }_{i}(\mathbf{x}_{o},t-t_{o}))\Theta
_{i}(\mathbf{r,\mathbf{\xi }}(t),t),  \notag
\end{eqnarray}%
\emph{where }$\Theta _{i}(r,\mathbf{\mathbf{\xi }}(t),t)$\emph{\ is the
strong occupation number (\ref{occupation number})}$,$ \emph{with} $\mathbf{%
\mathbf{\xi }}(t)\equiv \left\{ \mathbf{\mathbf{r}}_{1}(\mathbf{x}%
_{o},t-t_{o}),..,\mathbf{\mathbf{r}}_{N}(\mathbf{x}_{o},t-t_{o})\right\} $%
\emph{;}

\emph{B) in terms of }$f_{N}(\mathbf{x,}t)$ \emph{the 1-particle probability
density reads identically \ in }$\Gamma _{1}\times I$ \emph{as :}

\begin{equation}
f_{1}^{(N)}(\mathbf{y},t)=\int\limits_{\Gamma _{N}}d\mathbf{x}f_{N}(\mathbf{%
x,}t)\frac{1}{N}\sum_{i=1,N}\delta (\mathbf{y-x}_{i})\Theta _{i}^{\ast }(%
\mathbf{r,}t),  \label{one-particle PDF}
\end{equation}%
\emph{where} $\Theta _{i}^{\ast }(\mathbf{r,\mathbf{\xi }},t)$%
\begin{eqnarray}
\Theta _{i}^{\ast }(\mathbf{r,\mathbf{\xi ,}}t) &\equiv &1-\sum_{\substack{ %
j=1,N \\ i\neq j}}\Theta (d-\left\vert \mathbf{r-r}_{j}\right\vert )- \\
&&-\Theta (d/2-\left\vert \mathbf{r-r}_{W}\right\vert ),  \notag
\end{eqnarray}%
\emph{and } $\mathbf{\mathbf{\xi }}\equiv \left\{ \mathbf{\mathbf{r}}_{1},..,%
\mathbf{\mathbf{r}}_{N}\right\} .$ \emph{.}

\emph{Proof }

The proof follows by noting:

A) first, that Eqs. (\ref{one-particle PFD-dependence-initial-condition})
and (\ref{one-particle PDF}) mutually imply each other thanks to the
validity of Liouville equation for $f_{N}(\mathbf{x,}t)$ [see Eq.(\ref%
{integral Liouville eq}) in the Appendix] in the sub-domain of $\Gamma _{N}$
where no interactions (unary or binary) occur;

B) second, from Eq.(\ref{one-particle PDF}) there follows, in particular,
for $f_{1}^{(N)}(\mathbf{y},t)$ the identity
\begin{equation}
f_{1}^{(N)}(\mathbf{y},t)=\int\limits_{\Gamma _{2}}d\mathbf{x}_{1}d\mathbf{x}%
_{2}f_{2}(\mathbf{x}_{1},\mathbf{x}_{2}\mathbf{,}t)\delta (\mathbf{y-\mathbf{%
x}_{1}})\Theta _{1}^{\ast }(\mathbf{r,\mathbf{\xi ,}}t),  \label{f1-1}
\end{equation}%
which immediately implies Eq.(\ref{eq. for f1^N}) for $s=1.$ \textbf{c.v.d.}

\section{3 - A representation of $f_{1}^{(N)}(\mathbf{y},t)$ based on
functional continuation}

We notice that the proof of THM.1 can also be reached by introducing a
functional continuation - denoted as $f_{N}^{\ast }(\mathbf{x,}t)$ - of $%
f_{N}(\mathbf{x,}t)$ in the open subset ($\Gamma _{N}^{\ast \ast }$) of the
forbidden sub-domain $\Gamma _{N}^{\ast }$, where the particles of $S_{N}$
are not interacting with the boundary $\partial \Omega .$ The only minimal
requirement to be imposed on $f_{N}^{\ast }(\mathbf{x,}t)$ is that it
results continuous on the boundary set between $\Gamma _{N}^{\ast \ast }$
and $\overline{\Gamma }_{N},$ $\delta \Gamma _{N}^{\ast \ast }\cap \delta
\overline{\Gamma }_{N}$ (A). However, due to the freedom in its definition
it is always possible to require also that:

\begin{itemize}
\item B) $f_{N}^{\ast }(\mathbf{x,}t)$ is non-negative in the whole set $%
\Gamma _{N}\times I$ and strictly positive in $\Gamma _{N}^{\ast }\times I;$

\item C) $f_{N}^{\ast }(\mathbf{x,}t)$ is invariant with respect to
arbitrary permutations of like particles;

\item D) in the forbidden sub-domain $\Gamma _{N}^{\ast }$ $\ f_{N}^{\ast }(%
\mathbf{x,}t)$ satisfies the differential Liouville equation%
\begin{equation}
\frac{\partial }{\partial t}f_{N}^{\ast }(\mathbf{x,}t)+\sum\limits_{i=1,N}%
\mathbf{v}_{i}\cdot \nabla _{i}f_{N}^{\ast }(\mathbf{x,}t)=0.
\label{Liouville}
\end{equation}
\end{itemize}

It is obvious that Eq. (\ref{one-particle PFD-dependence-initial-condition})
remains valid even if $f_{N}(\mathbf{x,}t)$ is replaced by the an arbitrary
functional continuation satisfying these assumptions (A-D). This permits us
to reach the following integral representation for $f_{1}^{(N)}(\mathbf{y},t)
$:

\textbf{Corollary 1 of Thm.2 - Integral representation for }$f_{1}^{(N)}(%
\mathbf{y},t)$

\emph{In terms of the functional continuation} $f_{N}^{\ast }(\mathbf{x,}t)$
\emph{there results identically in} \emph{\ }$\Gamma _{1}\times I$ \emph{%
(for any }$\mathbf{y\equiv (r,v)\in \Gamma }_{1}$ \emph{and} $t\in
I\subseteq
\mathbb{R}
$\emph{):}%
\begin{equation}
f_{1}^{(N)}(\mathbf{y},t)=I_{1}^{(N)}-I_{2}^{(N)},
\label{representation-final}
\end{equation}%
\emph{where }$I_{1}^{(N)},I_{2}^{(N)}$ \emph{are the phase-space integrals}%
\begin{equation}
I_{1}^{(N)}=\int\limits_{\Gamma _{1(2)}}d\mathbf{x}_{2}f_{2}^{\ast (N)}(%
\mathbf{y},\mathbf{x}_{2}\mathbf{,}t)  \label{I1}
\end{equation}%
\emph{\ }%
\begin{equation}
I_{2}^{(N)}\equiv (N-1)\int\limits_{\Gamma _{1(2)}}d\mathbf{x}%
_{2}f_{2}^{\ast (N)}(\mathbf{y},\mathbf{x}_{2}\mathbf{,}t)\Theta
(d-\left\vert \mathbf{r-r}_{2}\right\vert )  \label{I2}
\end{equation}%
\emph{and the notation}
\begin{equation}
f_{2}^{\ast (N)}(\mathbf{y},\mathbf{x}_{2}\mathbf{,}t)\equiv
\int\limits_{\Gamma _{N}}d\mathbf{x}f_{N}^{\ast }(\mathbf{x,}t)\delta (%
\mathbf{y-x}_{1})  \label{f2-star}
\end{equation}%
\emph{has been introduced.}

\emph{Proof}

In fact in terms of $f_{N}^{\ast }(\mathbf{x,}t)$\ Eq.(\ref{one-particle PDF}%
) reads\bigskip

\begin{eqnarray}
&&\left. f_{1}^{(N)}(\mathbf{y},t)=\int\limits_{\Gamma _{N}}d\mathbf{x}%
f_{N}^{\ast }(\mathbf{x,}t)\frac{1}{N}\sum_{i=1,N}\delta (\mathbf{y-x}%
_{i})-\right.  \\
&&-\int\limits_{\Gamma _{N}}d\mathbf{x}f_{N}^{\ast }(\mathbf{x,}t)\frac{1}{N}%
\sum_{i=1,N}\delta (\mathbf{y-x}_{i})\sum_{\substack{ j=1,N \\ i\neq j}}%
\Theta (d-\left\vert \mathbf{r-r}_{j}\right\vert )  \notag
\end{eqnarray}

which delivers, upon imposing condition B,
\begin{eqnarray}
&&\left. f_{1}^{(N)}(\mathbf{y},t)=\int\limits_{\Gamma _{N}}d\mathbf{x}%
f_{N}^{\ast }(\mathbf{x,}t)\delta (\mathbf{y-x}_{1})-\right.  \\
&&-(N-1)\int\limits_{\Gamma _{N}}d\mathbf{x}f_{N}^{\ast }(\mathbf{x,}%
t)\delta (\mathbf{y-x}_{1})\Theta (d-\left\vert \mathbf{r-r}_{2}\right\vert
).  \notag
\end{eqnarray}

This equation reduces to (\ref{representation-final}) by introducing the
notation given above [see Eq.(\ref{f2-star})]. c.v.d.

We remark, furthermore that the following additional proposition holds:

\textbf{Corollary 2 (THM.2) }-\emph{\ }\textbf{Inequality for }$I_{2}^{(N)}$

\emph{In validity of THM.2, the phase-space integral }$I_{1}^{(N)}$ \emph{%
satisfies for all }$(\mathbf{r},\mathbf{v},t)\in \Gamma _{1}\times I$ \emph{%
the homogeneous equation }%
\begin{equation}
{F}_{1}(\mathbf{r},\mathbf{v},t)I_{1}^{(N)}=0,  \label{homogeneous equation}
\end{equation}%
\emph{where }$k_{\sup }$\emph{\ is a suitable strictly positive real
constant independent of }$N$\emph{. }

\emph{Proof.}

The proof is immediate thanks to Eq.(\ref{Liouville}) which, by assumption,
is satisfied by $f_{N}^{\ast }(\mathbf{x,}t).$

\section{4 - The strong B-G limit for $S_{N}$}

Let us now show how theorem 2 permits us to determine the strong B-G limit
of $f_{1}^{(N)},$ $f_{1}(\mathbf{y},t)\equiv L^{\ast }f_{1}^{(N)}.$ Here $%
L^{\ast }$ denotes the \emph{strong B-G limit operator} which is defined
\emph{in the sense of local convergence for ordinary functions defined in
phase-space} and is obtained letting $N\rightarrow \infty $ while requiring $%
d=c/N^{1/2},$ with $c$ a non-vanishing finite constant independent of $N$. \
In the sequel the limit operator $L^{\ast }$ acts on $f_{1}^{(N)}$ or $%
f_{2}^{(N)}$, both considered defined point-wise in suitable domains. \ In
particular, the $L^{\ast }$ coincides with the ordinary limit operator when
acting on an arbitrary real function of the parameters $N$ and $d.$ Let us
now assume that both $f_{1}^{(N)}(\mathbf{y},t)$ and $f_{1}^{\ast (N)}(%
\mathbf{y},t)\equiv \int\limits_{\Gamma _{1(2)}}d\mathbf{x}_{2}f_{2}^{\ast
(N)}(\mathbf{y,x}_{2},t),$ as well as the corresponding limit functions $%
f_{1}(\mathbf{y},t)\equiv L^{\ast }f_{1}^{(N)}(\mathbf{y},t)$ and $%
f_{1}^{\ast }(\mathbf{y},t)\equiv L^{\ast }f_{1}^{\ast (N)}(\mathbf{y},t)$
are bounded, i.e., $\ \sup (f_{1}),\sup (f_{1}^{\ast })<\infty $.

In such a case the following Lemma holds:

\bigskip

\textbf{Lemma (to THM.3) }-\emph{\ }\textbf{Inequality and B-G limit for }$%
I_{2}^{(N)}$

\emph{In validity of THM.2, let us assume that at least in a finite time
interval }$I_{o1}=\left[ t_{o},t_{1}\right] \subseteq I$ \emph{and} \emph{in
}$\Gamma _{2}^{\ast \ast },$ $f_{2}^{\ast (N)}(\mathbf{y},\mathbf{x}_{2}%
\mathbf{,}t)$ \emph{can be defined so that everywhere in }$\Gamma _{1}\times
I_{o1}$\emph{\ :}

\emph{a) the limit functions }$f_{s}\equiv L^{\ast }f_{s}^{(N)}$ \emph{and} $%
f_{s}^{\ast }\equiv L^{\ast }f_{s}^{\ast (N)}$ \emph{exist for }$s=1,2$
\emph{at least in }$\Gamma _{1}\times I_{o1};$

\emph{b)} $f_{s}^{(N)}$\emph{\ and }$f_{s}^{\ast (N)}$\emph{\ (for arbitrary
}$N\in
\mathbb{N}
$\emph{) as well as }$f_{s}$ \emph{and} $f_{s}^{\ast }$\emph{\ (for }$s=1,2$
$\emph{)}$ \emph{are, bounded ordinary functions defined at least in }$%
\Gamma _{1}\times I_{o1}$\emph{;}

\emph{c)}\ \emph{the phase-space integrals : }$\int\limits_{\Gamma _{1(2)}}d%
\mathbf{x}_{2}f_{2}^{\ast (N)}(\mathbf{y},\mathbf{x}_{2}\mathbf{,}t)\Theta
(d-\left\vert \mathbf{r-r}_{2}\right\vert )$ $\emph{\ }$\emph{(}$\emph{for\
any\ }N\in
\mathbb{N}
$\emph{)} \emph{and }$L^{\ast }\int\limits_{\Gamma _{1(2)}}d\mathbf{x}%
_{2}f_{2}^{\ast (N)}(\mathbf{y},\mathbf{x}_{2}\mathbf{,}t)\Theta
(d-\left\vert \mathbf{r-r}_{2}\right\vert )$ \emph{are bounded}$;$

\emph{d) the functions: }$f_{1}^{\ast (N)}(\mathbf{y,}t)$ \emph{\ (for any }$%
N\in
\mathbb{N}
$\emph{)} \emph{and }$f_{1}^{\ast }(\mathbf{y,}t)\equiv L^{\ast }f_{1}^{\ast
(N)}(\mathbf{y,}t)\equiv L^{\ast }\int\limits_{\Gamma _{1(2)}}d\mathbf{x}%
_{2}f_{2}^{\ast (N)}(\mathbf{y,x}_{2},t)$ \emph{are strictly positive,}

\emph{Then it follows that in } $\Gamma _{1}\times I_{o1}:$

\emph{L-I)} \emph{for any finite }$N\in
\mathbb{N}
$ \emph{the phase-space integral }$I_{2}^{(N)}$ \emph{can be majorized as
follows }%
\begin{equation}
I_{2}^{(N)}\leq (N-1)\frac{d^{3}}{V}k_{\sup }f_{1}^{(N)}(\mathbf{y},t),
\label{majorization}
\end{equation}%
\emph{where }$k_{\sup }$\emph{\ is a suitable strictly positive real
constant independent of }$N;$

\emph{L-II) uniformly in }$\Gamma _{1}\times I_{o1}$\emph{\ there results}$:$%
\emph{\ }%
\begin{equation}
L^{\ast }I_{2}^{(N)}=0.  \label{Lemma}
\end{equation}

\emph{Proof}

L-I) The proof is immediate. In fact, due to assumptions a) and b), together
with the strict positivity of $f_{1}^{\ast (N)}(\mathbf{y,}t)$ (assumption d)%
$,$ we can always require that in a finite time interval $I_{o1}=\left[
t_{o},t_{1}\right] \subseteq I$ \ there results
\begin{eqnarray}
&&\left. \int\limits_{\Gamma _{1(2)}}d\mathbf{x}_{2}f_{2}^{\ast (N)}(\mathbf{%
y},\mathbf{x}_{2}\mathbf{,}t)\Theta (d-\left\vert \mathbf{r-r}%
_{2}\right\vert )\leq \right.  \\
&\leq &\frac{d^{3}}{V}k_{\sup }\int\limits_{\Gamma _{1(2)}}d\mathbf{x}%
_{2}f_{2}^{\ast (N)}(\mathbf{y},\mathbf{x}_{2}\mathbf{,}t)=\frac{d^{3}}{V}%
k_{\sup }f_{1}^{\ast (N)}(\mathbf{y},t),  \notag
\end{eqnarray}%
where $k_{\sup }$ is a suitable strictly positive real constant. In
particular, since the inequality must hold for any $N>1,$ $k_{\sup }$ can
always be chosen as independent of $N$. \ This implies the inequality (\ref%
{majorization}).

L-II) To prove Eq.(\ref{Lemma}) let us invoke the majorization (\ref%
{majorization}) which implies
\begin{equation}
L^{\ast }I_{2}^{(N)}\leq L^{\ast }\left\{ (N-1)\frac{d^{3}}{V}k_{\sup
}f_{1}^{(N)}(\mathbf{y},t)\right\} ,
\end{equation}%
Due to assumption e) $\sup \left( f_{1}^{\ast }(\mathbf{y},t)\right)
<+\infty $ while $k_{\sup }$ is independent of $N.$ It follows \
\begin{equation}
L^{\ast }I_{2}^{(N)}\leq k_{\sup }\sup \left( f_{1}(\mathbf{y},t)\right)
L^{\ast }\left\{ (N-1)d^{3}\right\} .
\end{equation}%
Hence, since in the B-G limit by definition $L^{\ast }\left\{
(N-1)d^{3}\right\} =0$, this means that $L^{\ast }I_{2}^{(N)}$ is
identically zero in the set $\Gamma _{1}\times I_{o1}$. \textbf{c.v.d.}

We remark that to satisfy the condition of strict positivity here imposed on
$f_{1}^{\ast (N)}(\mathbf{y,}t)$ and\emph{\ }$f_{1}^{\ast }(\mathbf{y,}t)$
[see assumption d) in the Lemma] it is actually sufficient to require that $%
f_{1}^{(N)}(\mathbf{y,}t)$ and\emph{\ }its limit function\emph{\ }$f_{1}(%
\mathbf{y,}t)$ are strictly positive in $\Gamma _{1}\times I_{o1}.$ This is
because by definition $f_{1}^{\ast (N)}(\mathbf{y,}t)\geq $ $f_{1}^{(N)}(%
\mathbf{y,}t),$ while one can prove that $f_{1}^{\ast }(\mathbf{y,}t)=$ $%
f_{1}(\mathbf{y,}t)$ (see below). Then the following theorem has the flavor
of:

\bigskip

\textbf{Theorem 3 - Strong B-G limit for }$S_{N}$

\emph{In validity of THM.2 and the Lemma, assuming that the limit functions }%
$f_{1}(\mathbf{y},t)$\emph{\ and }$f_{1}^{\ast }(\mathbf{y},t)$\emph{\ exist
and are bounded at least in the space }$\Gamma _{1}\times I_{o1},$\emph{\
there it follows for }$S_{N}$\emph{\ that:}

\emph{T-1)\ uniformly in }$\Gamma _{1}\times I_{o1},$\emph{\ the strong B-G
limit function }$f_{1}(\mathbf{y},t)\equiv L^{\ast }f_{1}^{(N)}(\mathbf{y}%
,t) $\emph{\ reads }%
\begin{equation}
f_{1}(\mathbf{y},t)=f_{1}^{\ast }(\mathbf{y},t);  \label{eq-1}
\end{equation}

\emph{T-1) \ }$f_{1}(\mathbf{y},t)$\emph{\ satisfies identically the
homogeneous equation}%
\begin{equation}
{F}_{1}(\mathbf{r},\mathbf{v},t)f_{1}(\mathbf{y},t)=0.  \label{eq-2}
\end{equation}

\emph{Proof}

T-1) Second, thanks the Lemma [see Eq.(\ref{Lemma})], it follows that Eq.(%
\ref{representation-final}) of Corollary 1 delivers in the whole space $%
\Gamma _{1}\times I_{o1}$%
\begin{equation}
L^{\ast }f_{1}^{(N)}(\mathbf{y},t)=\int\limits_{\Gamma _{1(2)}}d\mathbf{x}%
_{2}L^{\ast }f_{2}^{\ast (N)}(\mathbf{y},\mathbf{x}_{2}\mathbf{,}t)\equiv
L^{\ast }f_{1}^{\ast (N)}(\mathbf{y},t),
\end{equation}%
which, denoting $f_{1}^{\ast }(\mathbf{y},t)=L^{\ast }f_{1}^{\ast (N)}(%
\mathbf{y},t),$ proves also Eq.(\ref{eq-1}).

T-2) Third, thanks to the Lemma, Eq.(\ref{eq-1}) delivers Eq.(\ref{eq-2}).
\textbf{c.v.d.}.

As consequence, we conclude that \emph{the strong B-G limit function }$f_{1}(%
\mathbf{y},t)$\emph{\ does not generally satisfy the limit equation (\ref%
{limit equation}) of the BBGKY (or Boltzmann) hierarchy}, i.e., in other
words \emph{the strong B-G limit does not exist for the equations of the
BBGKY hierarchy}.

It is manifest that this result can also be expressed in the following
equivalent form:

\bigskip

\textbf{Corollary of THM.3 }

\emph{In validity of THM.3 it follows that}
\begin{equation}
{\ }\left[ {{L^{\ast },F}_{{\ 1}}}\right] {{\ }f}_{1}^{(N)}(\mathbf{x}%
_{1},t)\neq 0.  \label{breaking of commutation rule}
\end{equation}

Namely the operators ${{L^{\ast }}}$ and ${{F}_{{\ 1}}}$ \emph{do not commute%
}. This means that for $S_{N}$ when applying the B-G limit operator ${{%
L^{\ast }}}$ to the equations of the BBGKY hierarchy \emph{the limit
equations do not generally recover the Boltzmann hierarchy. Hence, we
conclude that - in the case of the hard-sphere system here considered - the
Lanford conjecture for the BBGKY hierarchy fails, at least if it is intended
in the sense of the strong B-G limit here considered.}

\section{5 - Conclusions}

\ The main conclusion of this paper is that the Boltzmann hierarchy and the
Boltzmann equation \emph{cannot generally be recovered from the BBGKY
hierarchy for the system }($S_{N}$)\emph{\ of smooth-hard (and impenetrable)
spheres, at least in the sense of the strong B-G limit}. The discovery
appears striking, especially in view of the previous literature appeared on
the subject. A related interesting question which obviously arises, in the
light of this conclusion, is that of the interpretation of customary results
due to Lanford and followers. As is well-known, being all based on the
treatment of the smooth-hard sphere problem in terms of formal time-series
solution of the BBGKY, they rely on the interpretation of the B-G limit as a
weak$^{\ast }$ limit. This problem is which is particularly relevant in
connection with the issue of (local or global) validity of the Boltzmann
kinetic equation for the one-particle limit PSPD will be treated elsewhere%
\cite{Tessarotto2008}.

In this paper the validity of the Lanford conjecture \cite%
{Lanford1975a,Lanford1976,Lanford1981}, with particular reference to the
BBGKY hierarchy, has been investigated adopting, for this purpose, a $\Gamma
_{1}-$phase-space microscopic statistical description for the time-evolution
of a system of $N$ smooth-hard spheres ($S_{N}$). \ The result has been
achieved using the Klimontovich approach to MSD for $S_{N}.$ The approach,
which can in principle be extended to higher-order joint probability
densities, permits to determine a formal exact solution of their time
evolution without recurring to cumbersome time-series representations. This
allows, in particular, to construct an explicit integral representation for
the one-particle probability density, to be expressed in terms of the
initial $\Gamma _{N}$-phase-space microscopic probability density. \ The key
aspect of the approach here developed is - however - the fact that, since
the Klimontovich representation dose not involve the adoption of time-series
representations for the one-particle probability density usually adopted in
the customary BBGKY-approaches, it can be used to determine explicitly the
strong B-G limit of the joint probability densities. In this paper, in
particular, the behavior of the one-particle PSPD has been investigated.

We have shown that the one-particle limit function $f_{1}(\mathbf{y},t),$ in
the sense of the strong B-G limit, does not generally belong to the
functional class of the solutions of the corresponding limit equation (\ref%
{limit equation}).\ To reach the proof suitable assumptions on the behavior
of the one-particle and the two-particle joint probability densities - as
well as for the corresponding limit functions and related quantities - have
been invoked. This includes the hypothesis that the one-particle probability
density and its limit function are a suitably smooth and bounded ordinary
functions. \ Similar conclusions are expected to apply for arbitrary $s-$%
particle limit functions (for $s>1$).

\section*{Acknowledgments}

Useful comments by G. Tironi, Department of Mathematics and Informatics,
Trieste University are acknowledged. This work has been developed in
cooperation with the CMFD Team, Consortium for Magnetofluid Dynamics
(Trieste University, Trieste, Italy), within the framework of the MIUR
(Italian Ministry of University and Research) PRIN Programme: \textit{%
Modelli della teoria cinetica matematica nello studio dei sistemi complessi
nelle scienze applicate}. Support is acknowledged from GNFM (National Group
of Mathematical Physics) of INDAM (Italian National Institute for Advanced
Mathematics).

\section{Appendix - $\Gamma _{N}$-\textbf{MSD for} $S_{N}$}

The following basic result is well-known \cite%
{Grad1958,Cercignani1969,Cercignani1975}:

\bigskip

\textbf{Theorem 1 - }$\Gamma _{N}$-\textbf{MSD for} $S_{N}$

\emph{Let us assume that for }$S_{N}$\emph{\ (system of }$N$\emph{\ like
smooth-hard spheres) there results:}

\emph{I) \ for any finite }$N>0,$ \emph{the\ initial} \emph{microscopic
probability density} $f_{N}(\mathbf{x}_{o}\mathbf{,}t_{o})$ \emph{is defined
as an ordinary function which is at least differentiable in the whole }$6N$%
\emph{-dimensional phase space} $\Gamma _{N}=\Omega ^{N}\times
\mathbb{R}
^{3N}$\emph{; }

\emph{II)} $f_{N}(\mathbf{x}_{o}\mathbf{,}t_{o})$ \emph{is assumed assumed
suitably summable so that for any }$s=1,N-1$\emph{\ the initial }$s$\emph{%
-particle joint probability densities }
\begin{equation}
f_{s}^{(N)}(\mathbf{x}_{1}\mathbf{,..x}_{s},t_{o})=\int\limits_{\Gamma
_{1(s+1)}}dx_{s+1}f_{s+1}^{(N)}(\mathbf{x}_{1}\mathbf{,..x}_{s+1},t_{o})
\end{equation}%
\emph{\ are defined and suitably smooth in the }$s-$\emph{particle
phase-space }$\Gamma _{s}=\Omega ^{s}\times
\mathbb{R}
^{3s}.$

\emph{Then it always possible to define uniquely }$f_{N}(\mathbf{x,}t)$
\emph{in whole set }$\Gamma _{N}\times I,$ \emph{where }$I\equiv
\mathbb{R}
$\emph{\ (global existence domain)} \emph{such that:}

\emph{A) }$f_{N}(\mathbf{x,}t)$ \emph{is an ordinary smooth function} \emph{%
in whole set }$\Gamma _{N}\times I;$ \emph{\ }

\emph{B) in any open subset\ of }$I$\emph{\ not containing collision events }%
$f_{N}(\mathbf{x}(t)\mathbf{,}t)$\emph{\ is a continuous and differentiable
function of time which is constant on all} $\Gamma _{N}$\emph{-} \emph{%
phase-space trajectories }$\mathbf{x}(t)=\chi (\mathbf{x}_{o},t-t_{o}),$%
\emph{\ i.e. there results, at arbitrary }$t,t_{o}$ \emph{belonging to such
a time interval and arbitrary admissible initial state }$\mathbf{x}_{o}$
\emph{belonging to the subset of} $\Gamma _{N}$ \emph{\ in which no
collision event occur,}

\begin{equation}
f_{N}(\mathbf{x}(t)\mathbf{,}t)=f_{N}(\mathbf{x}_{o}\mathbf{,}t_{o})
\label{integral Liouville eq}
\end{equation}%
\emph{(integral Liouville equation); }

\emph{C) at an arbitrary collision time }$t_{c}\in I,$\emph{\ \ }$f_{N}(%
\mathbf{x}(t)\mathbf{,}t)$\emph{\ satisfies one of the following boundary
conditions:}

\emph{- unary collision:}
\begin{eqnarray}
&&\left. f_{N}(\mathbf{x}_{1}(t_{c})\mathbf{,..,\mathbf{x}_{i}^{(-)}(}t_{c}%
\mathbf{),..x}_{N}(t_{c}),t_{c})=\right.   \label{B-C-unary} \\
&=&f_{N}(\mathbf{x}_{1}(t_{c})\mathbf{,..,\mathbf{x}_{i}^{(+)}(}t_{c}\mathbf{%
),..x}_{N}(t_{c}),t_{c});  \notag
\end{eqnarray}

\emph{-binary collisions between particles }$i$\emph{\ and }$j:$

\begin{eqnarray}
&&\left. f_{N}(\mathbf{x}_{1}(t_{c})\mathbf{,..,\mathbf{x}_{i}^{(-)}(}t_{c}%
\mathbf{),\mathbf{x}_{j}^{(-)}}(t_{c})\mathbf{..x}_{N}(t_{c}),t_{c})=\right.
\\
&=&f_{N}(\mathbf{x}_{1}(t_{c})\mathbf{,..,\mathbf{x}_{i}^{(+)}(}t_{c}\mathbf{%
),\mathbf{\mathbf{x}_{j}^{(+)}}}(t_{c})\mathbf{,..x}_{N}(t_{c}),t_{c}).
\notag
\end{eqnarray}

\emph{D)} \emph{for each} $s=1,N-1$ \emph{the }$s$\emph{-particle joint
probability density }
\begin{equation}
f_{s}^{(N)}(\mathbf{x}_{1}\mathbf{,..x}_{s},t)=\int\limits_{\Gamma
_{1(s+1)}}dx_{s+1}f_{s+1}^{(N)}(\mathbf{x}_{1}\mathbf{,..x}_{s+1},t)
\end{equation}%
\emph{\ is uniquely defined for all }$t\in
\mathbb{R}
$ \emph{and satisfies the equation of the BBGKY hierarchy for }$%
f_{s}^{(N)};. $

\emph{R)} \emph{for }$s=2,N-1$ \emph{and} \emph{in any open subset\ of }$I$%
\emph{\ not containing collision events involving only the first }$s$ \emph{%
particles of} $S_{N},$%
\begin{eqnarray}
&&\left. f_{s}^{(N)}(\mathbf{x}_{1}(t)\mathbf{,..x}_{s}(t),t)=\right.  \\
&=&\int\limits_{\Gamma _{1(s+1)}}dx_{s+1}f_{s+1}^{(N)}(\mathbf{x}_{1}(t)%
\mathbf{,..\mathbf{..x}}_{s}(t)\mathbf{,x}_{s+1},t)  \notag
\end{eqnarray}%
\emph{\ is a continuous and differentiable function of time. (Proof omitted).%
}

\end{document}